Pedro Baptista de Castro[A,B]*, Kensei Terashima[A]*, Takafumi D. Yamamoto[A], Suguru Iwasaki[C], Ryo Matsumoto[A], Shintaro Adachi[A], Yoshito Saito[A,B], Hiroyuki Takeya[A] and Yoshihiko Takano[A,B]

[A]National Institute for Materials Science, 1-2-1 Sengen, Tsukuba, Ibaraki 305-0047, Japan

[B]University of Tsukuba, 1-1-1 Tennodai, Tsukuba, Ibaraki 305-8577, Japan

[C]Hokkaido University, Laboratory of Nanostructured Functional Materials, Research Institute for Electronic Science (RIES), N20 W10, Kita-ku, Sapporo, Hokkaido 001-0020, Japan

Corresponding authors:

Pedro Baptista de Castro E-mail: CASTRO.Pedro@nims.go.jp
National Institute for Materials Science, 1-2-1 Sengen, Tsukuba, Ibaraki 305-0047, Japan

Kensei Terashima E-mail: TERASHIMA.Kensei@nims.go.jp
National Institute for Materials Science, 1-2-1 Sengen, Tsukuba, Ibaraki 305-0047, Japan

# Effect of Dy substitution in the giant magnetocaloric properties of $HoB_2$

Recently, a massive magnetocaloric effect near the liquefaction temperature of hydrogen has been reported in the ferromagnetic material $HoB_2$. Here we investigate the effects of Dy substitution in the magnetocaloric properties of $Ho_{1-x}Dy_xB_2$ alloys ($x$ = 0, 0.3, 0.5, 0.7, 1.0). We find that the Curie temperature ($T_C$) gradually increases upon Dy substitution, while the magnitude of the magnetic entropy change $|\Delta S_M|$ at $T = T_C$ decreases from 0.35 to 0.15 J cm$^{-3}$ K$^{-1}$ for a field change of 5 T. Due to the presence of two magnetic transitions in these alloys, despite the change in the peak magnitude of $|\Delta S_M|$, the refrigerant capacity (*RC*) and refrigerant cooling power (*RCP*) remains almost constant in all doping range, which as large as 5.5 J cm$^{-3}$ and 7.0 J cm$^{-3}$ for a field change of 5 T. These results imply that this series of alloys could be an exciting candidate for magnetic refrigeration in the temperature range between 10 – 50 K.

## Introduction

Magnetic refrigeration is an emerging environmental friendly technology for refrigeration applications, as it does not require to use greenhouse gases and does not depend on conventional gas compression cycles[1–3] while having possible higher cycle efficiency[1,4]. It is based on the magnetocaloric effect (MCE), which consists of the adiabatic temperature change ($\Delta T_{ad}$) a magnetic material will undergo when a magnetic field is applied/removed adiabatically, but it can also be evaluated in terms of the magnetic entropy change ($\Delta S_M$) this magnetic material will undergo for the same field change, where $\Delta S_M$ usually peaks at the magnetic transition temperature ($T_{mag}$).

Recently, our group unveiled a giant magnetocaloric effect of $|\Delta S_M^{MAX}|$ = 0.35 J cm$^{-3}$ K$^{-1}$ (40.1 J kg$^{-1}$ K$^{-1}$) in the vicinity of a ferromagnetic transition at the Curie temperature ($T_C$) of 15 K for a field change of $\mu_0\Delta H$= 5 T in $HoB_2$[5]. Due to the closeness of its $T_C$ to the liquefaction point of hydrogen (20.3 K), this material became an attractive candidate for use in low-temperature magnetic refrigeration applications focused on the liquefaction stage of hydrogen. Hydrogen is considered to be one most promising replacement for hydrocarbon fuels as a clean energy source[6,7] and in particular liquid hydrogen is widely needed in the space industry[8] and its liquid form is one the suitable way for transportation and storage[9]. In this context, the discovery of magnetic materials with a high MCE effect at low temperatures is imperative for the development of such refrigerators working at cryogenic temperatures. Since the magnetocaloric effect peaks at $T_{mag}$, tuning the $T_C$ of $HoB_2$ to a higher temperature is of extreme interest to examine $HoB_2$-based materials as possible candidates for refrigeration before the liquefaction stage, especially below temperatures of 77 K. Since, $DyB_2$ orders ferromagnetically at $T_C$ = 50 K[10,11] and exhibits a $|\Delta S_M|$ of 0.16 J cm$^{-3}$ K$^{-1}$ (17.1 J kg$^{-1}$ K$^{-1}$) for $\mu_0\Delta H$= 5 T[12], the partial substitution of Ho by Dy is expected to shift $T_C$ to higher values in the expense of a probable reduction of $|\Delta S_M|$. In this work, we study the magnetocaloric properties of $Ho_{1-x}Dy_xB_2$ alloys ($x$ = 0, 0.3, 0.5, 0.7, 1.0) and compare with other well-known materials working at the same temperature span.

All the magnetocaloric properties of the samples will be reported in volumetric units (J cm$^{-3}$ K$^{-1}$) as this is the adequate unit when comparing materials for application purposes as there is a volume limit when constructing real applications[13,14].

Therefore, herein all comparisons with other materials will be done in this unit by converting it using the ideal density of each material when not provided.

## Experimental Section

### *Sample Synthesis*

Polycrystalline samples of $Ho_{1-x}Dy_xB_2$ were prepared by an arc-melting process in a water-cooled copper hearth arc furnace under Ar atmosphere. Stoichiometric amounts of Ho (99.9% purity), Dy (99.9% purity), and B (99.5% purity) were weighted and then arc melted several times. During the process of sample synthesis, we found out that annealing the samples under different conditions did not change the X-ray diffraction patterns of the obtained samples, therefore no annealing procedure was carried out in the samples in this work.

### *Characterization*

Powder X-ray diffraction (XRD) patterns of the arc melted samples were investigated using a Rigaku-MiniFlex 600 with Cu K$\alpha$ radiation. The lattice parameters, the volume of the unit cell, and density were obtained by refining the XRD patterns using the FULLPROF[15] software.

### *Magnetization Measurements*

Magnetization measurements were carried out by a superconducting quantum interference device magnetometer contained in the Magnetic Property Measurement System XL (Quantum Design). Zero-field cooling (ZFC) and field cooling (FC) measurement at low fields were taken to evaluate the evolution of $T_C$ as a function of Dy content. For the evaluation of $|\Delta S_M|$ the magnetization measurements of the sample under various applied fields ranging from 0.01 to 5 T were performed in ZFC process.

## Results and Discussion

### *Crystal Structure*

Fig. 1 (a) shows the XRD patterns for the obtained arc melted samples. The main phase peaks can be indexed into a hexagonal *P*6/*mmm* $AlB_2$ type crystal structure as shown by the red fitting curves. The remaining peaks are assigned as $REB_4$, unreacted RE or $RE_2O_3$ (RE = Ho, Dy) impurity peaks marked by a black square (■), a black star (★), or a black diamond (♦) respectively. The obtained lattice parameters, the volume of the unit cell, and density are summarized in Table 1.

As shown in Table 1, Dy substitution in the Ho site seems to strongly affect the *c*-axis length while the *a*-axis length weakly changes, illustrated in Fig. 1 (b) where we plot the normalized lattice parameters ($a/a_0$ and $c/c_0$) by the value of $x = 0$. Both $c/c_0$ and $a/a_0$ increases with $x$, roughly following the so-called Vegard's law (marked by the dashed black line), but with different rates. The observed changes in the lattice constants in $Ho_{1-x}Dy_xB_2$ suggests that the substitution of Ho by Dy in the $REB_2$ main phase was successful, and these partially substituted samples can be in the form of a random alloy. We note that in the case of $HoB_{2-x}Si_x$ solid solutions[10] where B site is partially substituted, it has been reported that the expansion rate of *a*-axis length and *c*-axis length are comparable to each other. This difference in the change of lattice constants between $Ho_{1-x}Dy_xB_2$ and $HoB_{2-x}Si_x$ implies that the *a*-axis and *c*-axis lengths in $HoB_2$-based compounds might be closely related to the bonds along axes. Namely, *c*- axis length seems to be depending on Ho-B bonds and be sensitive to both rare-earth and B-site atoms, while the *a*-axis length might be more dependent on the B site atom.

### *Magnetic Properties*

The ZFC-FC isofield magnetization (*M-T*) for an applied field of $\mu_o H$ = 0.01 T and isothermal magnetization (*M-H*) at $T$ = 5 K are shown in Fig. 2 (a-e) and (f-g) for each obtained sample respectively. For the Dy containing samples, the divergence between the ZFC and FC *M-T* curves becomes more pronounced and a small magnetic hysteresis in the *M-H* curves is observed, including the end-material $DyB_2$.

To evaluate the of the magnetic transition temperatures in this system, the temperature-dependent derivative of the ZFC curves were taken and are shown in the lower panels of Fig. 2 (a-e). The Curie temperatures, that are defined by the peak position in ∂M/∂T curves marked by the $T_C$ arrows, showing a systematic increase with Dy content. On the other hand, a second magnetic transition marked by $T^*$ that is observed at lower temperatures, which is also observed at $HoB_2$ at $T^*$ = 11 K[5] and $DyB_2$ at $T^*$ = 15 K[12], seems to be almost unchanged by partial substitution of Dy. The origin of $T^*$ was attributed to a possible spin-reorientation mechanism[12], however, the nature of this transition is still unknown and its investigation is outside the scope of this work. The Dy doping dependence of both transitions is summarized in Fig. 3 showing the monotonic increase of $T_C$ until 50 K, while $T^*$ remains almost constant.

### *Magnetocaloric Properties*

For evaluating the magnetocaloric effect of the obtained samples, *M-T* curves in a wide range of applied magnetic fields were measured for all samples, shown in Fig. 4 (a-e) and $|\Delta S_M|$ was calculated using the Maxwell relation:

$$\Delta S_M = \mu_0 \int_0^H \left(\frac{\partial M}{\partial T}\right)_H dH \;\; (1).$$

The obtained $|\Delta S_M|$ for fields ranging from 0 to 5 T is shown in Fig. 4 (f-g).

Due to the presence of the two transitions at $T^*$ and $T_C$, two peaks appear at $|\Delta S_M|$. Therefore, here we will define and compare the maximum entropy change $|\Delta S_M^{MAX}|$ as $|\Delta S_M(T = T_C)|$, since $T^*$ remains almost unchanged during the whole doping range and is always lower than 15 K while we are interested in the $|\Delta S_M|$ peak shifted toward higher temperature by Dy doping. In this way, the obtained values of $|\Delta S_M^{MAX}|$ for $\mu_0\Delta H$ = 5 T were 0.35, 0.3, 0.18, 0.16 and 0.15 J cm$^{-3}$ K$^{-1}$ for $x$ = 0, 0.3, 0.5, 0.7 and 1.0 respectively.

In addition to the change in the magnitude of $|\Delta S_M^{MAX}|$, an interesting characteristic appears in the $|\Delta S_M|$ curves of $Ho_{1-x}Dy_xB_2$. That is, since there are multiple transitions in this series of alloys, even though there is a net loss at $|\Delta S_M^{MAX}|$, the entropy change curve shows an increase of $\delta T_{FWHM}$, defined as the region in the entropy curve where $|\Delta S_M| \geq |\Delta S_M^{MAX}|/2$, leading to a gain in maximum entropy change for higher temperature spans. Such a widening of the $|\Delta S_M|$ curves due to multiple transitions has been commonly observed in materials that show more than one magnetic transition [16–18] and it tends to lead to a high figure of merits. The $|\Delta S_M|$ for all samples for a field change of $\mu_0\Delta H$ = 5 T is shown in Fig. 5.

Here we consider the two most commonly used figures of merit for evaluating magnetocaloric materials, (i) the refrigerant capacity[1,2,19] ($RC$) defined as $RC = -\int_{T_{cold}}^{T_{hot}} \Delta S_M dT$ , where the usual values of $T_{cold}$ and $T_{hot}$ are the ones defined by $\delta T_{FWHM}$ (as such, $\delta T_{FWHM} = T_{hot} - T_{cold}$), that tries to represent the amount of heat that can be transferred from the cold sink to the hot sink in an ideal thermodynamic cycle, and (ii) the relative cooling power[1,20–22] ($RCP$): $RCP\ (S) = \Delta S_M^{MAX} * \delta T_{FWHM}$. For a field change of $\mu_0\Delta H$ = 5 T, the obtained values of $RC$ and $RCP$ on $Ho_{1-x}Dy_xB_2$ are summarized in Table 2 where it clearly shows that due to the increase of $\delta T_{FWHM}$, they stay almost the same for the whole doping range, indicating that the loss in maximum $|\Delta S_M|$ is compensated by the increase of $\delta T_{FWHM}$.

Let us compare the magnetocaloric property values of $|\Delta S_M^{MAX}|$, *RC*, *RCP*, and $T_{mag}$ in $Ho_{1-x}Dy_xB_2$ with those of representative materials that are often considered for magnetic refrigeration applications, such as $RAl_2$ (R=rare-earth) series and materials that exhibit high values of $|\Delta S_M|$ with transition temperatures ranging up to 77 K, based on Fig. S5 of Ref [5]. For this purpose, the values of entropy change are converted into volumetric units by using the density contained in the AtomWork[23] database, unless otherwise provided by the authors. Also, the values of *RC*/*RCP* and $\delta T_{FWHM}$ are estimated from the reported within the contained references when not reported by the authors and they are all summarized in Table 2.

At the temperature range of 15- 20 K, $HoB_2$ and $Ho_{0.7}Dy_{0.3}B_2$ show a higher figure of merit by a large margin, for $\mu_0\Delta H$ = 5 T, when compared to compounds with similar $T_{mag}$ such as $ErAl_2$[24], TmGa[25], EuS[26], HoN[27] and $DyNi_2$[24]. For the materials with transition temperatures around 30 K, $Ho_{0.5}Dy_{0.5}B_2$ also shows a significantly larger figure of merit than the compounds with similar $T_{mag}$ such as $HoAl_2$[28] and $ErCo_2$[29]. On the other hand, $Ho_{0.3}Dy_{0.7}B_2$ shows similar performance to the intermetallic compound HoNi[30], but both of them surpass the $ErCo_2$-based alloy $Er_{0.53}Ho_{0.47}Co_2$[31]. In the case of $DyB_2$, even though it shows almost half of $|\Delta S_M|$ compared to $Gd_3Ru$[32], the figure of merits in those two compounds are comparable to each other, and both vastly outperform those values in the Laves phase compound $DyAl_2$[33]. Note, that here these compounds were selected for comparison for exhibiting the highest $|\Delta S_M^{MAX}|$ with similar transition temperature to the alloys the Ho-Dy-B alloys, to the best of our knowledge, or for being already used in magnetic

refrigeration prototypes. In most of the mentioned cases, even when the $|\Delta S_M^{MAX}|$ of the Ho-Dy-B alloys is lower, their figure of merit is, higher by a larger margin. In addition, it should be noted that even among compounds with successive magnetic transitions with similar $|\Delta S_M^{MAX}|$ such as $Ho_2Cu_2Cd$[18], ErGa[34] and $Ho_2Au_2In$[35], $Ho_{1-x}Dy_xB_2$ show remarkably high *RC*/*RCP* values. (See Table 2).

Thus, these results indicate that $Ho_{1-x}Dy_xB_2$ alloys have a great potential for use in magnetic refrigeration ranging from 10 to 55 K when compared to the similar materials with $T_{mag}$ within this range, making these compounds as useful as other compounds with giant $|\Delta S_M|$ values in this temperature range such as $ErCo_2$ and $Gd_3Ru$.

## Conclusions

In this work, we evaluate systematically the effects of Dy substitution on the giant magnetocaloric effect of $HoB_2$. Even if there is a net loss in the peak value of the $|\Delta S_M|$, due to the increase of $\delta T_{FWHM}$ in these alloys, the *RC* and *RCP* remain extremely high compared to compounds working in a similar temperature range. Therefore, this series of alloys would have a high potential to work as magnetic refrigerants in the temperature range from 10-50 K.

**Acknowledgments**

We acknowledge fruitful discussions with Mohammed Elmassalami. This work was partly supported by the JST-Mirai Program “Development of advanced hydrogen liquefaction system by using magnetic refrigeration technology”, the JSPS KAKENHI, and the JST-CREST. P.B. Castro acknowledges the scholarship support from the Ministry of Education, Culture, Sports, Science and Technology (MEXT), Japan.

## Disclosure Statement

The authors declare no conflict of interest

**Funding**

This work was supported by the JST-Mirai Program (Grant No. JPMJMI18A3), JSPS KAKENHI (Grant Nos. 19H02177, 20K05070), JST CREST (Grant No. JPMJCR20Q4)

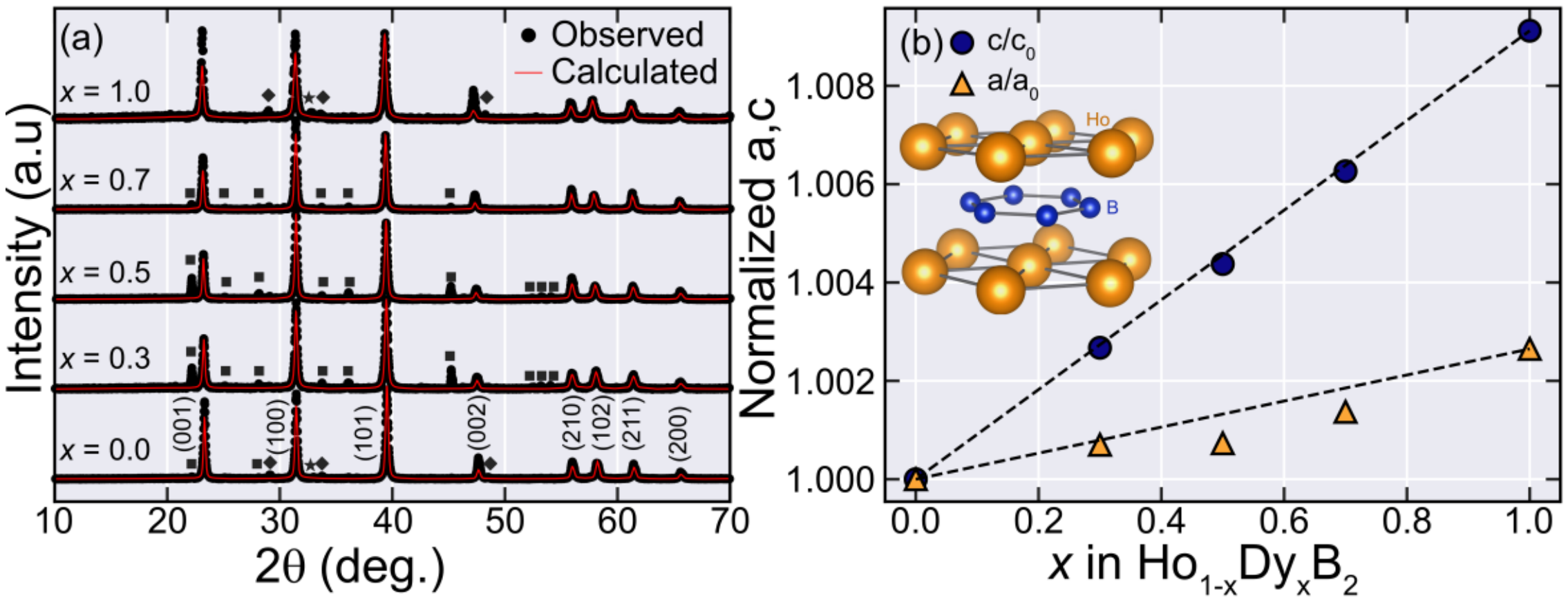

**Figure 1.** Powder XRD patterns and lattice constant evolution for $Ho_{1-x}Dy_xB_2$ alloys. (a) XRD patterns of the obtained alloys. The red lines show the calculated patterns from Rietveld refinement for the $REB_2$ main phase. The black square (■) marks an $REB_4$ impurity phase, while the black star (★) marks a RE impurity peak and the black diamond (♦) marks a $RE_2O_3$ impurity peak (RE=Ho, Dy). (b) The lattice parameters normalized by the value at $x$ = 0, as a function of $x$. The black dashed line shows a guide based on Vegard's law, given by $1 - x + x\left(\frac{a_1, c_1}{a_0, c_0}\right)$ where $(a,c)_{0,1}$ is the lattice constants at $x$ = 0 or 1.

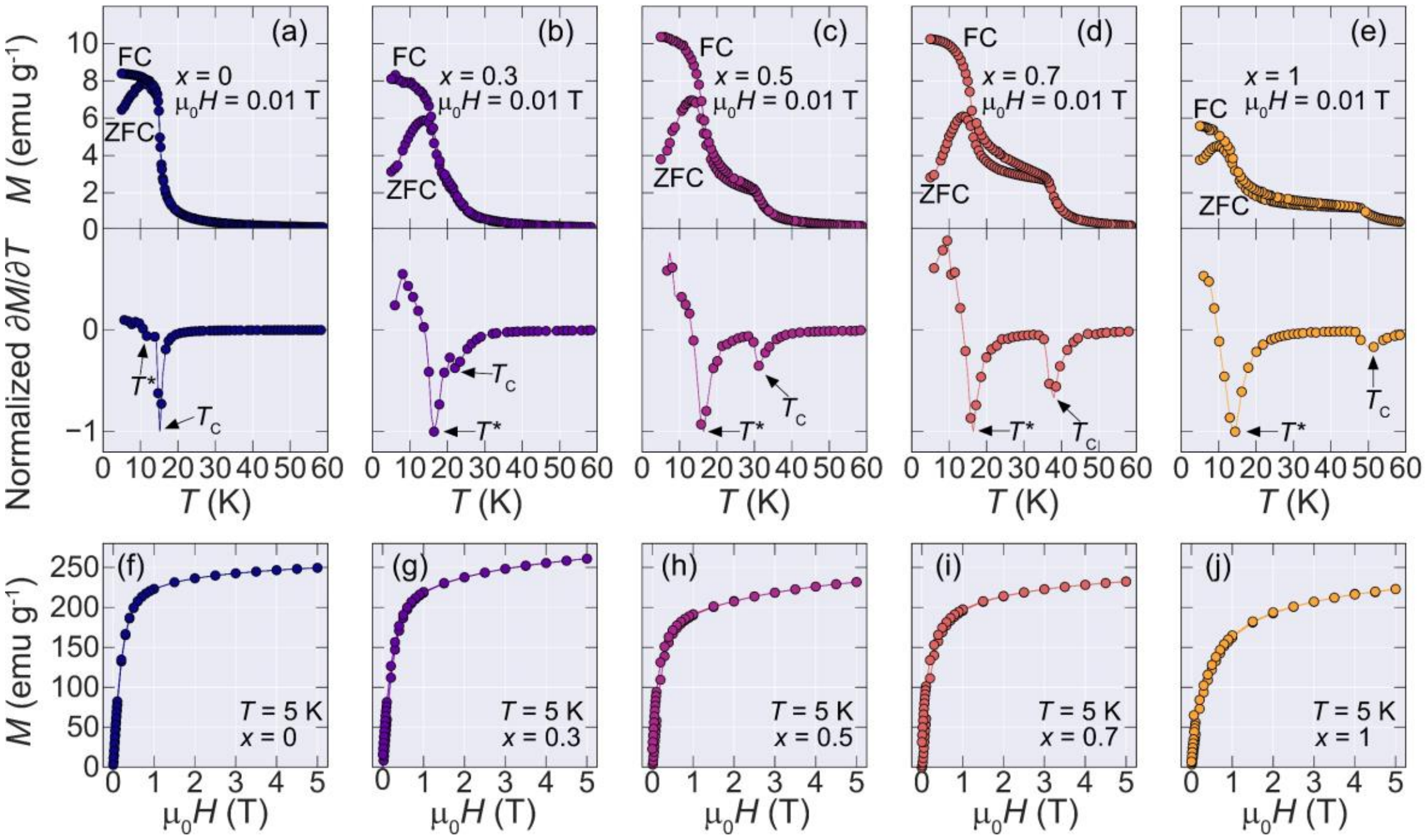

**Figure 2.** Isofield (*M-T)* ZFC-FC, Normalized temperature-dependent derivatives of the ZFC curves, and Isothermal (*M-H*) magnetization curves of $Ho_{1-x}Dy_xB_2$ alloys. (a-e) ZFC and FC curves for all synthesized alloys for an applied field of $\mu_0 H$ = 0.01 T. The lower panels show the derivatives of the ZFC curves normalized by the minimum of the derivative value. The two magnetic transitions $T_C$ and $T^*$, are marked by the arrows. (f-g) Isothermal magnetization at $T$ = 5 K.

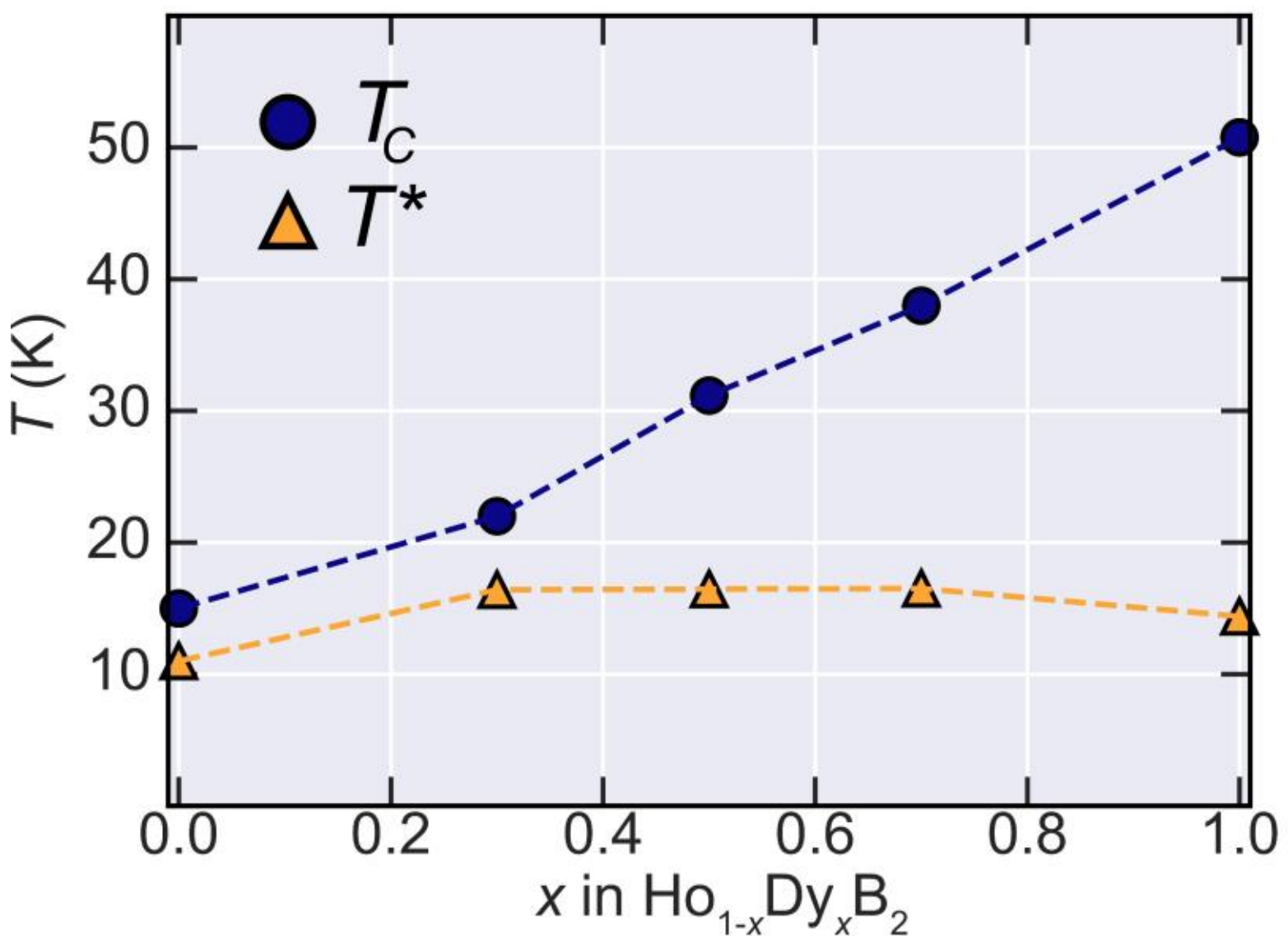

**Figure 3.** Phase diagram between ordering temperature and doping amount ($x$) for $Ho_{1-x}Dy_xB_2$. The blue filled circles show the evolution of $T_C$ while the yellow triangles show $T^*$. While $T_C$ increases monotonically with $x$, $T^*$ remains almost constant.

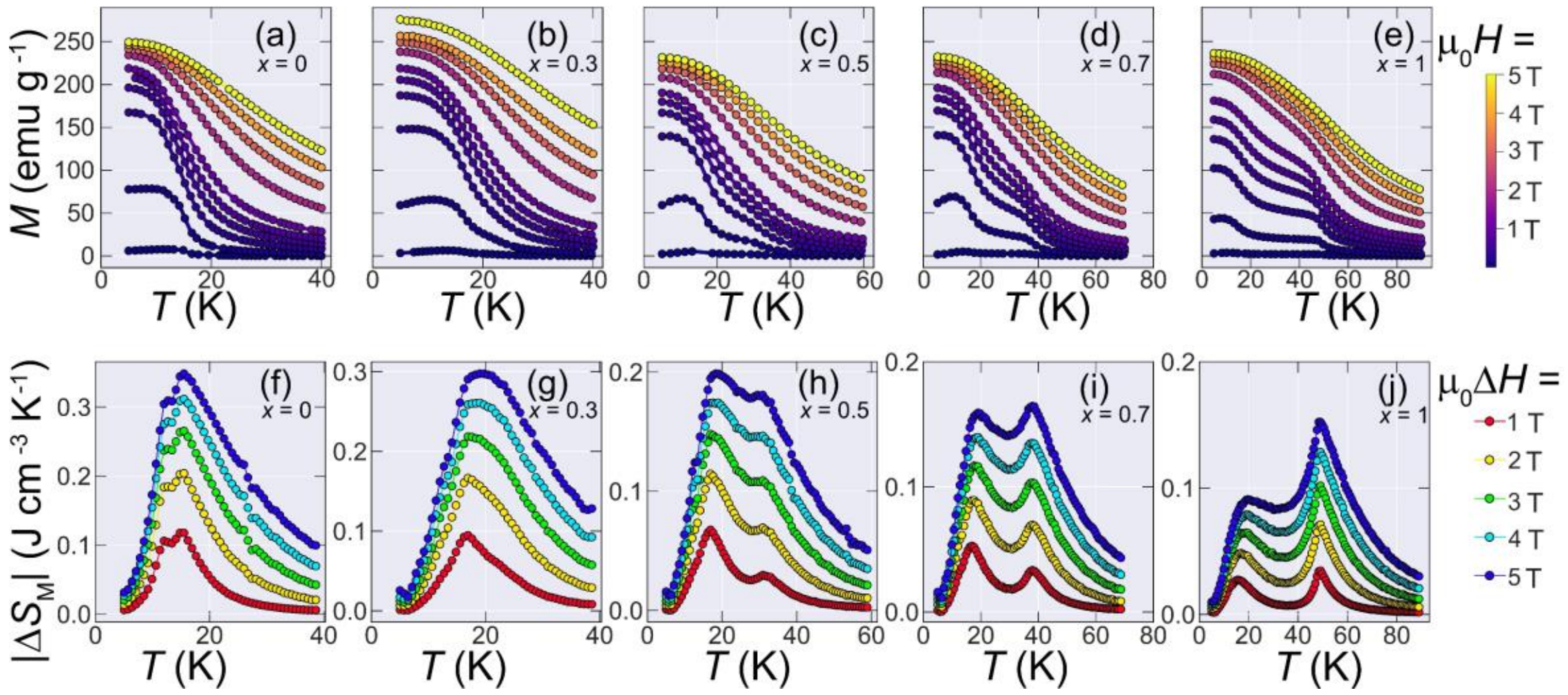

**Figure 4.** *M*-*T* curves at a vast range of applied fields and obtained magnetic entropy change for $Ho_{1-x}Dy_xB_2$ alloys. (a-e) The obtained M-T curves measured by ZFC process from $\mu_0 H$ = 5 T to 0.01 T. (f-j) Magnetic entropy changes for $Ho_{1-x}Dy_xB_2$ alloys for $\mu_0\Delta H$ ranging from 1 to 5 T obtained from the *M*-*T* curves of (a-e). With the increase of Dy content, the maximum value of $|\Delta S_M|$ decreases from 0.35 J cm$^{-3}$ K$^{-1}$ ($x$ = 0) to 0.16 J cm$^{-3}$ K$^{-1}$ ($x$ = 1.0).

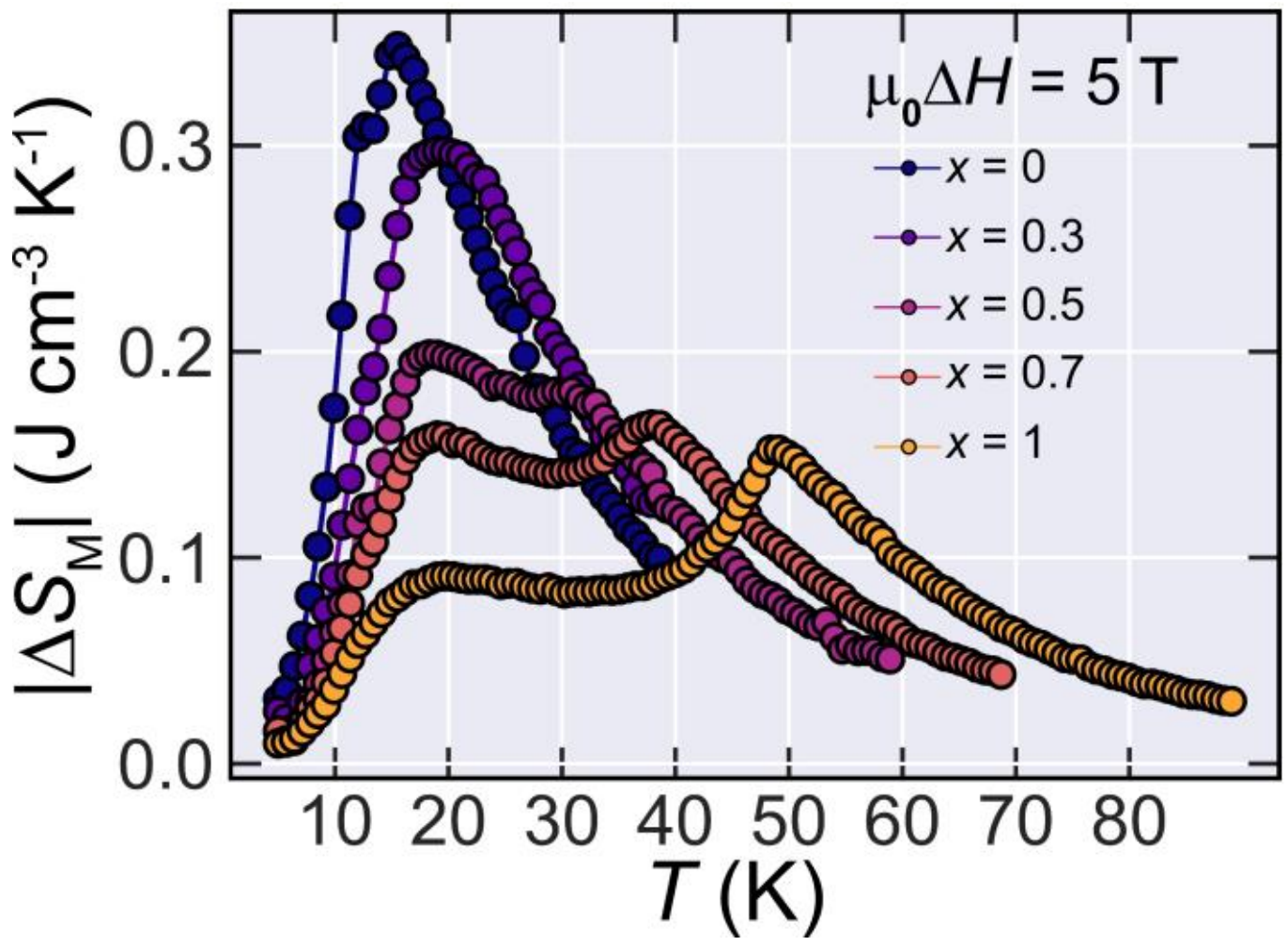

**Figure 5.** $|\Delta S_M|$ at $\mu_0\Delta H$ = 5 T for $Ho_{1-x}Dy_xB_2$ alloys.

| Nominal Dy (x) | a (Å) | c (Å) | V ($Å^3$) | ρ ($g/cm^3$) |
|---|---|---|---|---|
| 0 | 3.283(3) | 3.815(8) | 35.62(4) | 8.696 |
| 0.3 | 3.285(6) | 3.827(1) | 35.77(8) | 8.626 |
| 0.5 | 3.285(7) | 3.832(5) | 35.82(9) | 8.591 |
| 0.7 | 3.287(8) | 3.839(7) | 35.94(7) | 8.540 |
| 1.0 | 3.292(0) | 3.850(6) | 36.14(0) | 8.461 |

Table 1: Obtained lattice parameters from XRD patterns for $Ho_{1-x}Dy_xB_2$

| Material | $T_{mag}$ (K) | $\Delta|S_M^{MAX}|$ (J $cm^{-3}$ $K^{-1}$) | $\delta T_{FWHM}$ (K) | *RC* (J $cm^{-3}$) | *RCP* (J $cm^{-3}$) | Ref |
|---|---|---|---|---|---|---|
| $HoB_2$ | 15 | 0.35 | 18.9 | 5.1 | 6.6 | This work |
| $Ho_{0.7}Dy_{0.3}B_2$ | 22 | 0.30 | 23.5 | 5.5 | 7.0 | This work |
| $Ho_{0.5}Dy_{0.5}B_2$ | 31 | 0.18 | 32.8 | 5.5 | 6.3 | This work |
| $Ho_{0.3}Dy_{0.7}B_2$ | 37 | 0.16 | 42.7 | 5.8 | 7.0 | This work |
| $DyB_2$ | 50 | 0.15 | 51.2 | 5.1 | 7.8 | This work |
| $ErAl_2$ | 14 | 0.22[a] | 15.4[b] | 2.7[a,b] | 3.5[a,b] | 24 |
| TmGa | 15 | 0.3[a] | 14.6[b] | 3.2[a] | 4.4[a,b] | 25 |
| EuS | 18 | 0.21[a] | 20.9[b] | 3.4[b] | 4.5[a] | 26 |
| HoN | 18 | 0.29 | 19.9 | 4.4[b] | 5.7[b] | 27 |
| $DyNi_2$ | 21[b] | 0.24[a,b] | 21.4[b] | 3.8[a,b] | 5.0[a,b] | 24 |
| $Ho_2Au_2In$ | 21 | 0.16[a] | 26.7 | 3.0[a] | 4.2[a] | 35 |
| $Ho_2Cu_2Cd$ | 30 | 0.18[a] | 23.6[b] | 3.0[a] | 4.3[a] | 18 |
| ErGa | 30 | 0.18[a] | 0.18[a] | 4.2[a] | 5.6[a,b] | 34 |
| $HoAl_2$ | 30 | 0.15[a,b] | 31.8[b]- | 3.6[a,b] | 4.9[a,b] | 28 |
| $ErCo_2$ | 30 | 0.37[a,b] | 10.1[b] | 3.2[a,b] | 3.8[a,b] | 29 |
| HoNi | 36 | 0.17[a] | 42.3[b] | 5.2[a,b] | 7.3[a,b] | 30 |
| $Er_{0.53}Ho_{0.47}Co_2$ | 35[d] | 0.22[b] | 15.1[b] | 3.0[b] | 3.5[b] | 31 |
| $Gd_3Ru$ | 54 | 0.26 | 23.5[b] | 5.2[b] | 6.1 | 32 |
| $DyAl_2$ | 56 | 0.11[a,b] | 42.3[b] | 3.8[a,b] | 4.9[a,b] | 33 |

Table 2: Values of $|\Delta S_M^{MAX}|$, $\delta T_{FWHM}$, *RC* and *RCP* for $Ho_{1-x}Dy_xB_2$ and representative compounds with similar transitions temperatures. All the data is expressed in volumetric units. a: Converted to volumetric units using ideal density contained in the AtomWork database. b: Estimated from the $|\Delta S_M|$ curves reported in the reference.